\documentclass{aa} % for a paper on 1 column

\usepackage[varg]{txfonts}
\usepackage{subfigure}
\usepackage{caption}

\newcommand{\sii}{\ensuremath{\left[\mathrm{S~II}\right]}}
\newcommand{\pii}{\ensuremath{\left[\mathrm{P~II}\right]}}
\newcommand{\sx}{\ensuremath{\left[\mathrm{S~IX}\right]}}
\newcommand{\feiimilledeux}{\ensuremath{\left[\mathrm{Fe~II}\right]} 1257 }
\newcommand{\pagam}{Pa \ensuremath{\gamma}}
\newcommand{\pabet}{Pa \ensuremath{\beta}}
\newcommand{\six}{\ensuremath{\left[\mathrm{Si~X}\right]}}

\begin{document}

\title{New insights on the central stellar population and gas outflow in NGC 1068 from YJH spectroscopy with SPHERE/VLT}

\author{Pierre Vermot\inst{1}  \and Yann Clénet\inst{1}       \and Damien Gratadour\inst{1}}

\institute{LESIA, Observatoire de Paris, Université PSL, CNRS, Sorbonne Université, Université de Paris, 5 place Jules Janssen, 92195 Meudon, France}

\date{Received 17 April 2019 / Accepted 14 July 2019}

\abstract{\textit{ Aims :} We characterise the properties of stars, dust, and gas and their spatial distribution in the central region of the Seyfert 2 galaxy NGC1068. \\
\textit{Method :} Our study is based on near-infrared (YJH, 0.95-1.650 um, R=350) long-slit spectroscopy observations of the central region of NGC 1068 with a 0.4" spatial resolution. 
We decomposed the observed continuum emission into three components: hot dust, stars, and scattered light from the central engine. We measured their contributions at various distances from the nucleus. We also measured fluxes and Doppler shifts for the emission lines in our spectrum to probe the physical conditions of the narrow line region. \\ 
\textit{Results :} Dust and stars are the main sources of continuum emission, but scattered light from the central engine has also been detected in the very central region. Together, these three components reproduce the observed continuum well. The dust emission is compatible with a 830~K blackbody. It has only been detected in the very central region and is not spatially resolved. The stellar content is ubiquitous. It harbours a 250 pc cusp centred around the nucleus, over-imposed on a young stellar background. The spectrum of the cusp is consistent with a 120 Myr old single stellar population. Finally, the emission lines exhibit a significant Doppler shift that is consistent with a radial outflow from the nucleus in a biconical structure. The $\left[Fe\ II\right]$ behaviour strongly differs from other lines, indicating that it arises from a different structure.}

\keywords{Infrared: Galaxies; Galaxies: active, Seyfert, stellar content, individual: NGC 1068; Techniques: spectroscopic}
\titlerunning{New insights on the central stellar population and gas outflow in NGC 1068}
\authorrunning{P. Vermot et al.}
\maketitle 

\section{Introduction}

The spiral galaxy NGC 1068 ((R)SA(rs)b) has frequently been observed because of its proximity ($D_{A}=14.4~\mathrm{Mpc} \Leftrightarrow  70~\mathrm{pc}.\mathrm{arcsec}^{-1}$) and high luminosity \citep[several $ 10^{11}~\mathrm{L}_{\odot}$  according to][]{Pier1994}. This makes it central to understanding active galactic nuclei (AGN).  NGC 1068 is the archetypal Seyfert 2 galaxy: it harbours a distinct narrow line region (NLR) (Macchetto1994) and a strong dusty source of infrared radiation \citep{Gratadour2006} that hides a Seyfert 1 nucleus \citep{Antonucci1985, Antonucci1993}. However, the observations revealed many different structures in the nuclear region, with complex interactions and spatial distributions that still left questions about the physical processes that take place in AGN unanswered. \\
NGC 1068 harbours a supermassive back hole (SMBH) with a dynamically estimated mass of $8.6\ \pm 0.5 \times 10^6\ M_\odot$ \citep{Lodato2003} that is associated with an accretion disc, which is the main source of the luminosity in the commonly accepted description of AGN \citep{Rees1984, Hickox2018} and is referred to as central engine (CE). This CE is hidden from the observer by an obscuring material that is composed of gas and dust \citep{Hickox2018} and is often referred to as a \textit{\textup{dusty torus}}. This dusty structure is known to be on a  parsec scale \citep{Poncelet2006, Lopez-Gonzaga2014}, and its temperature ranges from 320 K \citep{Lopez-Gonzaga2014} to possibly up to 1200 K \citep{Gratadour2003}. In radio wavelengths, two radio jets that extend to several hundred parsecs from the SMBH are observed \citep{Wilson1983}. These jets seem to interact with the close environment of the nucleus: they are surrounded by several clouds that are detected in the mid-infrared \citep{Bock2000}, and the northern jet is redirected by at least one of them that is located at 25 pc north of the centre \citep{Gallimore1996}. \\
A conspicuous NLR with a size similar to that of the jets (several hundred parsec) was originally observed in the UV and optical wavelengths \citep{Macchetto1994, Kraemer2000, Cecil2002} and is also observed in the near-infrared \citep{Davies2007a,Martins2010a, Exposito2011, Riffel2014}. These observations revealed that the NLR has a biconical shape that is oriented NE-SW, with evidence of motion of the gas: the northern part is blueshifted and the southern part is redshifted. This has been interpreted as a radial outflow from the nucleus with an inclined biconical structure \citep{Das2006}. Two main ionisation mechanisms for the NLR are used to explain the behaviour of the observed emission lines: photoionisation from the CE \citep{Kraemer2000, Hashimoto2011}, and ionisation from shocks caused by interaction with the jet \citep{ Dopita1996,Kraemer2000, Exposito2011}. Stars have also been mentioned as a possible source of ionisation \citep{Nazarov1996, Exposito2011}. \\ 
As observed by \cite{Origlia1993} and confirmed by further studies, stars are undoubtedly present around the nucleus of NGC 1068. Stellar population synthesis   \citep{Martins2010} and emission line ratios diagnostics \citep{Origlia1993} both indicate a relatively recent star formation around the nucleus. \\
Even if these four components (CE, dust, NLR and stars) in the nucleus of NGC 1068 are clearly confirmed, many questions remain, for instance, about their spatial distribution,  their relative contribution to the flux at different wavelengths,  the physical characteristics of this region (dust temperature, gas kinematics, and ionisation mechanisms), and about the age of the stellar population. \\
We here investigate these questions using SPHERE at the \textit{Very Large Telescope}, and in particular, its long-slit spectroscopy mode.  Compared to similar instruments (SINFONI, X-SHOOTER), we selected SPHERE for the unique combination of high angular resolution ability, spectral range, and relatively wide field of view. The spectral range of this instrument gives us access to a spectral region of interest for our study (YJH, $ 0.95 - 1.65\ \mu m $). Firstly, it is an intermediate spectral domain where the thermal emission from hot dust, stars, and the CE can contribute to the continuum, making it feasible to study the three of them simultaneously. Secondly, it contains many emission lines, giving us the opportunity to investigate the physical properties of the gas and its excitation mechanisms. Finally, the extinction by dust, even if still present at these wavelengths, is much lower than in the optical domain, giving us the possibility to examine deeper regions. \\
Section 2 describes the observations and the data reduction. Section 3 presents our analyses of the continuum emission and spectral features, which are then discussed in Sect. 4. Section 5 is dedicated to a brief summary and conclusions.

\section{Observation and data reduction}

\subsection{Observations}

SPHERE \citep{Beuzit2008} is a facility with extreme adaptive optics (AO) system and coronagraph installed at the UT3 Nasmyth focus of the VLT. It is primarily designed for the characterisation of exoplanets. Observations were carried out in the night of December 5, 2014, during the science verification program. We used the IRDIS subsystem in its long-slit spectroscopy mode \citep[LSS,][]{Vigan2008} with medium resolution (MRS), covering the range $ 0.95 - 1.65\ \mathrm{\mu m} $ with $R \sim 350$. We used a classical Lyot coronagraph with 0.2" radius, which we first positioned on the bright core source to avoid contamination of the surrounding region (dataset called SPH1), then we positioned it with a slight offset in order to observe this bright core (SPH2). In both cases the spatial scale is $12.25\ \mathrm{mas.pixel}^{-1}$, the linear dispersion is $1.13\ \mathrm{nm.pixel}^{-1}$ (measured from the calibration lamp emission lines), and the slit (11" $\times$ 0.09") crossed the photocenter with $PA = 12^{\circ}$.\\ The sky emission was measured outside the galaxy, 144" south-east to the nucleus. These observations were followed by the observation of BD+00413, a close G0 calibration star with accurate 2MASS photometry in the H band, $H = 9.591 \pm 0.026$ \citep{Cutri2003}. The airmass was 1.1 at the beginning of the observation and 1.2 at the end.  The seeing was not good (from 1.2" to 1.5" according to the observation log), which led to a poor AO correction and a final spatial resolution of 0.35" in the H band \citep[FWHM measured on the central source, which is known to be almost unresolved with SPHERE in the H band,][]{Gratadour2015, Rouan2019}. Integration time and number of exposures are presented in Table~\ref{table_obs} for each dataset.

\begin{table}[ht]
\centering
\caption{\label{table_obs} Observation summary}
\begin{tabular}{c|c|c|c}
Observation & Number of & Duration of one & Integration \\
name  & exposures  & exposure (s) & time (s)\\
\hline 
&&&\\
SPH1 & 21 & 64 & 1344 \\
SPH2 & 15 & 64 & 960 \\
SKY & 9 & 64 & 576 \\
BD+00413 & 3 & 64 & 192 \\
\end{tabular}
\\
\end{table}

\subsection{Data reduction}
We divided the spectral images by the instrument flat-field response, which was computed with a per-pixel linear fit on lamp flat-field acquisitions with various integration times. A bad-pixel map was created by selecting pixels whose response or variability were outside a 3 $\sigma$ region around the associated mean on all pixels. Bad-pixel values were replaced by a linear interpolation of the surrounding pixels. We corrected the shear distortion of the images using the correction matrix provided by the ESO manual. We also corrected the second-order spectral distortion measured on the calibration lamp spectrum. Each individual SPHERE image contains two spectral images associated with the two polarimetric channels of the instrument. Because we did not use the polarimetric mode, both channels were isolated and treated as independent images. We finally performed the mean of all spectral images. We applied this method to the object, the calibration star, and the sky images. Finally, we combined SPH1 and SPH2 by taking the mean of the overlapping regions (accounting for a 0.4" spatial offset between the two observations) and performed sky subtraction. \\
We performed wavelength calibration with a linear fit on the positions of the calibration lamp emission lines. We then measured atmospheric transmission by comparing the spectrum of the reference star to the predicted spectrum of a G0 star \citep{Pickles1998}, and corrected the object spectra for this. We finally performed flux calibration with the 2MASS H-magnitude of BD-00413. \\
We associate the uncertainty on each spectral image pixel with the standard deviation of its values on the different frames before they were averaged. The flux calibration  is an important contributor to our errors ($\sim 30 \% $). \\

\begin{figure}[ht]
   \includegraphics[width=0.44\textwidth]{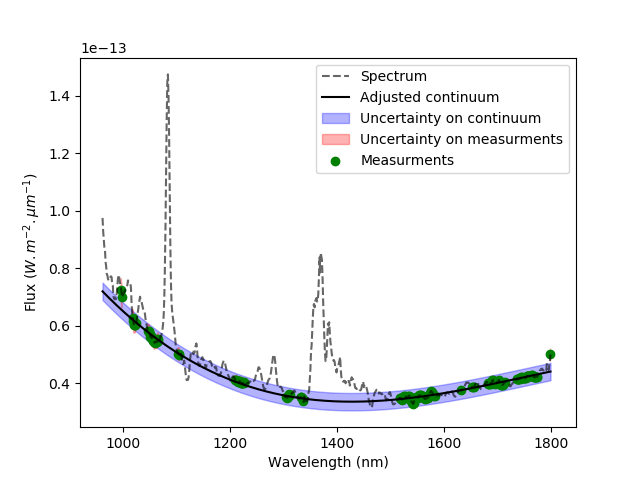}
   \caption{\label{conti_full} Extraction of the continuum for the spectrum integrated over the entire slit length. The dashed line represents the observed spectrum, the green disks represent the sample values for continuum measurement, and the solid line is the fitted fourth-degree polynomial. The uncertainty on the adjusted continuum is displayed in blue. The location of the emission lines is shown in green, and the location of absorption lines is shown in red (some cannot be shown in this figure because they are visible only in specific regions).}
\end{figure}

\begin{figure}[ht]
   \includegraphics[width=0.44\textwidth]{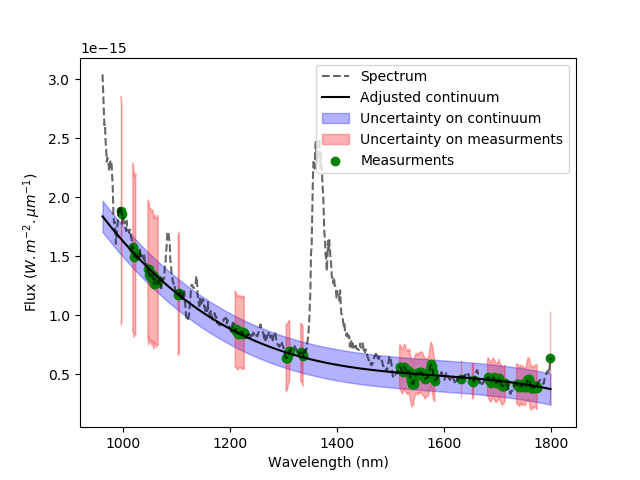}
   \caption{\label{conti_north} Same as Fig.~\ref{conti_full}, but for a 0.3" region 3" north of the nucleus. Uncertainties on the sample values are shown in red.}
\end{figure}

\begin{figure}[ht]
   \includegraphics[width=0.44\textwidth]{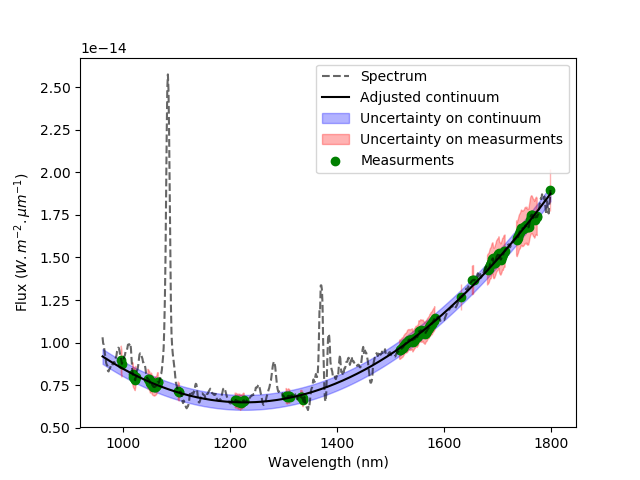}
   \caption{\label{conti_center} Same as Fig.~\ref{conti_full}, but for the central 0.3" region. Uncertainties on the sample values are shown in red.}
\end{figure}
\section{Analysis}

We constructed two synthetic spectral images, one solely containing the continuum emission, the other solely containing the spectral features. To estimate the continuum emission, we selected a few wavelengths that were free from emission or absorption features and fitted a fourth-degree polynomial to these points at each position of the slit. This method provided us with a continuum spectral image that we subtracted from the original in order to obtain the continuum-free contribution alone.

\subsection{Continuum emission analysis}

Examples of the continuum extraction for three regions are presented in Figs.  \ref{conti_full}, \ref{conti_north}, and \ref{conti_center}. Figure \ref{conti_full} presents the spectrum integrated on the entire spatial dimension and the continuum extracted from it. It exhibits a U-shape with a minimum located around 1400 nm. In the same way, Figure \ref{conti_north} represents the spectrum integrated over a 0.3" aperture 3" north of the nucleus, with the associated continuum emission. It shows an increase in flux for short wavelengths. The southern spectrum is very similar. Finally, Fig. \ref{conti_center} is the integrated spectrum of the central region with an aperture of 0.3". It  strongly emits towards long wavelength. \\
The resulting continuum spectral image shows a \textbf{significant} uniform background. Its contribution has been measured by taking the mean continuum on a 0.5" aperture at the extreme south of the slit (identical to the background measured at the extreme north of the slit) and was subtracted from the whole spectrum. We briefly discuss the nature of this continuum at the end of this section. Otherwise, the continuum is considered to be background subtracted.  \\ 

\begin{figure}[ht]
   \includegraphics[width=0.44\textwidth]{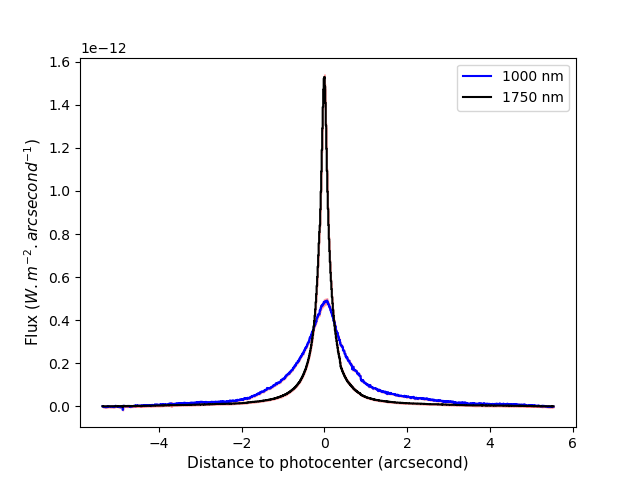}
   \caption{\label{profile} Spatial profile of the continuum emission at short ($\lambda=1000\ nm$, blue curve) and long ($\lambda=1750\ nm$, blue curve) wavelengths.}
\end{figure}

The profiles presented in Fig. \ref{profile} show that the 1750 nm component is unresolved with our spatial resolution.  The 1000 nm component has a much wider profile. \\

%\begin{figure}[ht]
%   \caption{\label{profile_log} Spatial profile of the continuum emission at short and long wavelengths, with logarithmic y-axis}
%   \includegraphics[scale=0.5]{profile_2_log.png}
%\end{figure}

The spectral behaviour, location, and spatial extent of these two main components give us strong indications regarding their nature. The component that peaks at long wavelengths is very likely related to the obscuring material that is predicted by the unified model of AGN. The other component that peaks at short wavelengths is associated with the stellar population. As mentioned before, studies also indicate that light from the CE may contribute significantly to the continuum \citep{Crenshaw2000,Martins2010a, Gratadour2015}. We therefore tried to identify it as well.\\

In this part, we distinguish the spatial contributions from the stars, hot dust, and the CE in the observed NIR continuum emission. For this purpose we simulated spectra for these three components and adjusted their relative contributions to the measured continuum. \\

The stellar continuum does not include any emission or absorption line. It is complex because it results from the superposition of stars at different temperatures, whose relative contributions to the flux depend on the age of the population. To represent the stellar component, we therefore used the single stellar population (SSP) evolutionary model provided by GALEV \citep{Kotulla2009} with a Salpeter initial mass function (IMF). Using this model, we extracted spectra for ages ranging from very young (t = 8 Myr) to intermediate (t = 400 Myr) stellar populations (intermediate age cannot be distinguished from an older population in our spectral range) with a 4 Myr time step (maximum resolution of the GALEV output). Considering that the circumnuclear region (defined hereafter as the region located farther away than 1" from the
photocenter) is mainly composed of stars, we used the continuum measured in this region to test
our populations. We conclude that the 120 Myr population provides the best fit (in the least-squares sense) to our data. \\

The CE component can be modelled by a power law $F_{\lambda} = \lambda^{-\alpha}$ typically with $0 < \alpha < 1.5$. According to the AGN unified model, a good estimate for the spectral shape of the CE can be obtained from Seyfert 1 spectra. In their AGN atlas, \cite{Riffel2006} observed that the near-infrared (NIR) spectra of Seyfert 1 galaxies can be well described by a broken power law with a steep continuum below 1100 nm and an almost flat continuum redwards. We assume that the blue steep continuum arises from the stars, and thus that the central region contribution is mainly flat. From now on we therefore use $\alpha = 0$ to represent the scattered contribution from the CE. Higher values of $\alpha$ do not change our result significantly: the slope of the associated spectrum remains well below the slope from the stellar component. \\

Dust is assumed to be present at temperatures ranging from 320 K \citep{Jaffe2004} to sublimation at possibly up to T$\sim$1800 K for graphite grains \citep{Mor2012}. When we assume the 120 Myr stellar population and the scattered light flat spectrum, the best fit in the central region is obtained with a 830 K blackbody. \\

\begin{figure}[ht]
   \includegraphics[width=0.44\textwidth]{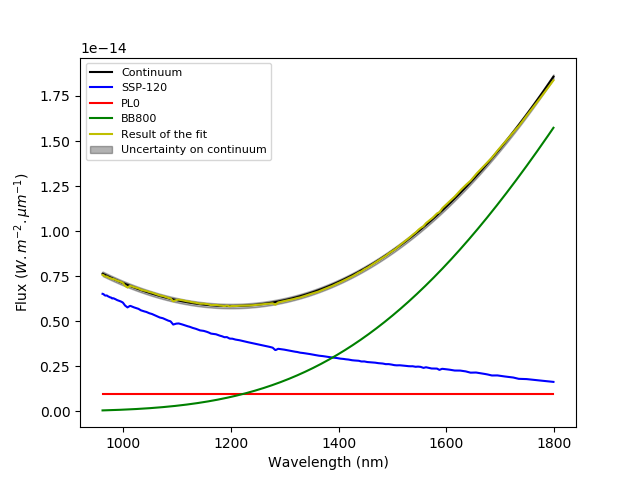}
   \caption{\label{fit_conti_dust} Adjustment of the continuum in the central 0.3". The black line represents the measured continuum. The blue, green, and red lines represent the adjusted contribution of the 120 Myr single stellar population (SSP-120 in the legend), the 830~K blackbody (BB830), and the flat contribution from scattered light (PL0). The yellow line represents the sum of these contributions.}
\end{figure}

\begin{figure}[ht]
   \includegraphics[width=0.44\textwidth]{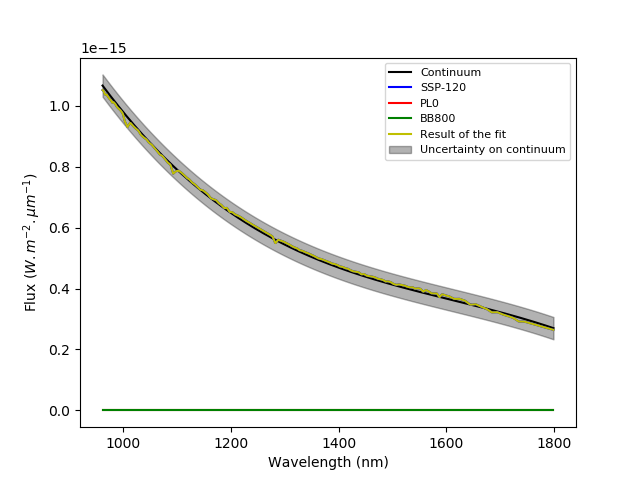}
   \caption{\label{fit_conti_stars} Same as Fig. \ref{fit_conti_dust}, but for a 1" aperture around the +2" north position. The fitting procedure indicates that stellar light is the only source of continuum emission, therefore the curves \textit{SSP-120} and \textit{Result of the fit}  overlap and cannot be identified separately. The same holds for the PL0 and BB830 contributions, which are nulls.} 
\end{figure}

The continuum emission can then be very well decomposed into three components: the 830~K blackbody, the 120~Myr single stellar population, and $\lambda^{-0}$ emission. Figure \ref{fit_conti_dust} displays the decomposition of the continuum integrated on the 0.3" central region. We can observe that the three components are significantly present even when the hot dust largely dominates. Figure \ref{fit_conti_stars} represents the same decomposition for a spectrum integrated from 2" to 2.5" north of the CE. This confirms that this region is largely dominated by stars. \\

\begin{figure}[ht]
   \includegraphics[width=0.44\textwidth]{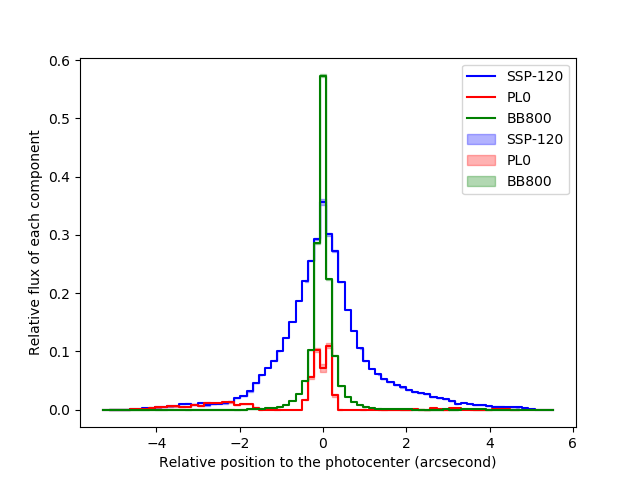}
   \caption{\label{plot_fluxs} Flux of each component relative to the total flux of the photocenter. The blue, green, and red lines represent the adjusted contribution of the 120 Myr single stellar population, the 830~K blackbody, and the flat continuum.}
\end{figure}

When it is performed at every position of the slit, this decomposition produces spatial profiles for the three components, which are presented in Fig. \ref{plot_fluxs}. This reveals an unresolved central component that is dominated by dust with a significant portion of light coming from the central nucleus. The stellar content is distributed in a cluster with FWHM $\sim$ 1.5" centred on the nucleus.
\\

%
%\begin{figure}[ht]
%   \includegraphics[width=0.44\textwidth]{chi2.png}
%   \caption{\label{chi2_map} Residuals map as a function of the dust temperature and the stellar population age. The residuals are expressed in $W.m^{-2}.\mu m^{-1}$.}
%\end{figure}

%\begin{figure}[ht]
%   \includegraphics[width=0.44\textwidth]{fit_conti_full.png}
%   \caption{\label{fit_conti_full} Same as Fig. \ref{fit_conti_dust} but for the continuum integrated on the entire slit length.} 
%\end{figure}

\subsubsection*{Continuum background analysis}

\begin{figure}[ht]
   \includegraphics[width=0.44\textwidth]{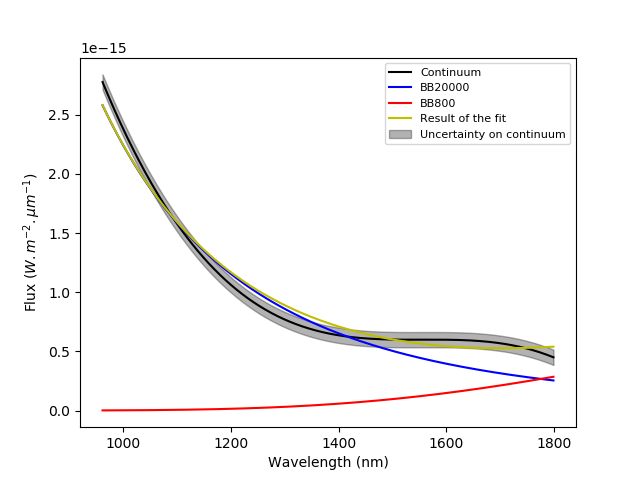}
   \caption{\label{fit_conti_back_good} Final adjustment of the background continuum. The black line represents the measured continuum. The blue and red lines represent the fitted contributions of the 20000 K and 830~K blackbodies, respectively. The yellow line represents the sum of these contributions.} 
\end{figure}

The components used to describe the central region fail to reproduce the observed background continuum. Changing the stellar population for a younger one can slightly improve the quality of the fit, but our best explanation (presented in Fig. \ref{fit_conti_back_good}) is that the background is largely dominated by very hot stars, with a little contribution from hot dust. Such stars in YJH can simply be modelled by a blackbody in a Rayleigh-Jeans regime (valid for T > 12000 K) and provide a good fit to our data. This important contribution from hot stars is surprising, but it is consistent with the very high starburst activity detected in NGC 1068 by previous studies \citep[][and references therein]{Romeo2016}. 

%\begin{figure}[ht]
%   \includegraphics[width=0.44\textwidth]{fit_conti_background_fail.png}
%   \caption{\label{fit_conti_back_fail} Same as Fig. \ref{fit_conti_dust} but for the background continuum.} 
%\end{figure}

\subsection{Analysis of spectral features}
This subsection describes the properties of the spectral features we detected in the continuum-free image that results from the decomposition described in the introduction of this section. At least 11 emission and 4 absorption features are detected. 

The emission lines were identified following \cite{Martins2010a}. We present this in Table~\ref{table_lines}. The identification of the absorption features is discussed in Sect. \ref{discussion}. \\

\begin{table}[ht]
\centering
\caption{\label{table_lines} Emission line summary. The wavelengths are given in rest frame. The measure of the spatial extent is approximate and is provided for qualitative comparisons.}
\begin{tabular}{c|c|c|c}
Identification & Wavelength & Maximum flux  & Spatial \\
of the line & (nm) & position (") & extent (") \\
\hline 
& & & \\
$\left[\mathrm{S~VIII}\right] $ & 991 & / & / \\
He~II & 1012 & 0.15 & 4.0 \\
$\left[\mathrm{S~II}\right] $ & 1032 & 0.3 & 3.5 \\
He~I & 1083 & 0.0 & 4.0 \\
Pa~$\gamma$ & 1093 &  0.0 & 4.0 \\
$\left[\mathrm{P~II}\right] $ & 1188 & 0.0 & 3.5 \\
$\left[\mathrm{S~IX}\right] $ & 1252 & 0.0 & 3.0 \\
$\left[\mathrm{Fe~II}\right] $ & 1257 & 0.9 & 8.0 \\
Pa~$\beta$ & 1272 & 0.0 & 4.0 \\
$\left[\mathrm{Si~X}\right] $ & 1430 & 0.2 & 2.5 \\
$\left[\mathrm{Fe~II}\right] $ & 1643 & 0.8 & 7.0 \\
\end{tabular}
\end{table}

\subsubsection*{Spatial distribution of the emission lines}
\label{spatial distri}
In this part we present our measurements of the emission line flux distribution along the slit. We used two methods to compute the fluxes of the emission lines. The first method is to compute the sum of the pixel values throughout the spectral range where flux is detected for the considered line. This method is robust and can be applied even in regions with a low signal-to-noise ratio. We therefore used it for most of our lines. A double-Gaussian fit was used as a second method for the two pairs of overlapping lines He I - Pa $\gamma$ and [S IX] - [Fe II] 1257. In order to improve the quality of our flux measurements for the [S IX] - [Fe II] 1257 overlapping lines, we set the central wavelengths of the lines as fixed parameters in our fit and imposed an equal width on both lines. We extracted spatial profiles for ten of our lines. Their main characteristics (position of the maximum emission and spatial extent) are presented in Table~\ref{table_lines}. We were unable to extract information on $\left[\mathrm{S~VIII} \right]$ because it is located in a region with low signal-to-noise ratio and may result from the contribution of different close emission lines that were detected in \cite{Martins2010}. The positions of the maximum emission and spatial extent highlight two categories of lines. The first category contains all the lines whose maximum emission is located near the photocenter and whose spatial extent is smaller than 4" ($\sim 350\ pc$). The majority of the lines falls in this category, with the exception of the two $\left[\mathrm{Fe~II}\right]$ lines, which form a second category whose maximum emission is located at 0.8" and 0.9" north, respectively and whose spatial extent is twice as large as those of the first category. \\

\subsubsection*{Doppler shift in emission lines}

Using a Gaussian fit on the emission lines with a high signal-to-noise ratio, we were able to measure shifts in their central wavelengths. These shifts are associated with a Doppler effect and are discussed in speed units. The rest speed is measured in the photocenter. \\

\begin{figure}[ht]
   \includegraphics[width=0.44\textwidth]{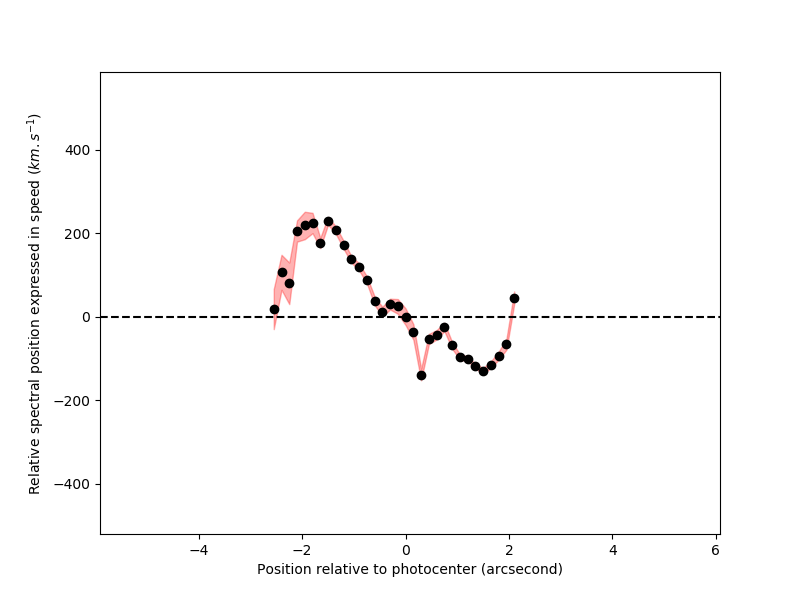}
   \caption{\label{he_i_speed} Doppler shift of $He\ I$, measured from 0.15" bins. The speed is positive for a redshift and is measured relative to its value at the photocenter. The red region represents 3 $\sigma$ uncertainties.}
\end{figure}

\begin{figure}[ht]
   \includegraphics[width=0.44\textwidth]{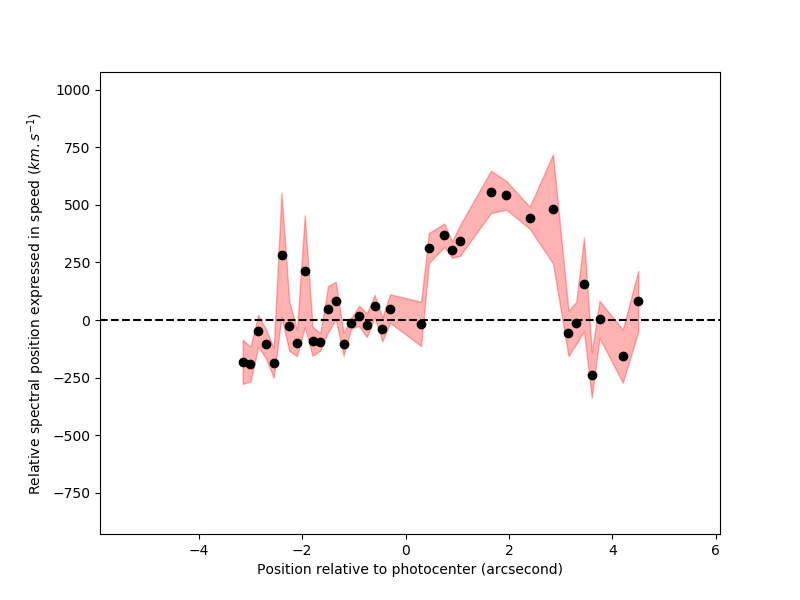}
   \caption{\label{fe_ii_speed} Doppler shift of $\left[\mathrm{Fe II}\right]$ 1643, measured from 0.15" bins. The speed is positive for a redshift and is measured relative to its value at the photocenter. The red region represents 3 $\sigma$ uncertainties.}
\end{figure}

The two categories we previously identified based on the spatial distribution are still relevant here: the behaviour of $\left[\mathrm{Fe~II}\right]$~1643 and other measurable lines differs strongly (no velocity shift can be measured for $\left[\mathrm{Fe~II}\right]$~1257 because it overlaps $\left[\mathrm{S~IX}\right] $). We present the results for He~I and $\left[\mathrm{Fe~II}\right]$~1643 in Figs. \ref{he_i_speed} and \ref{fe_ii_speed}, still considering that He~I is representative of the other emission lines. Velocity plots for other lines are presented in Appendix B when available. The two velocity profiles are very different in shape and amplitude.

Starting from the centre, He~I exhibits an increase in speed in both directions, blueshifted in the north and redshifted in the south, until it reaches a maximum just before 2" ($\sim 150\ pc$) where it quickly slows down to the rest speed. With respect to the CE location, the shift increases to $140 \pm 10\ km.s^{-1}$ in the northern region and up to $210 \pm 25\ km.s{-1}$ in the south.  
$\left[\mathrm{Fe~II}\right]$~1643, on the other hand, exhibits a redshift in both directions, a strong redshift the northern region ($600 \pm 100\ km.s^{-1}$) and a weaker redshift in the south ($100 \pm 50\ km.s^{-1}$).

\subsubsection*{Spatial profiles of the equivalent width of absorption lines}

\begin{figure}[ht]
   \includegraphics[width=0.44\textwidth]{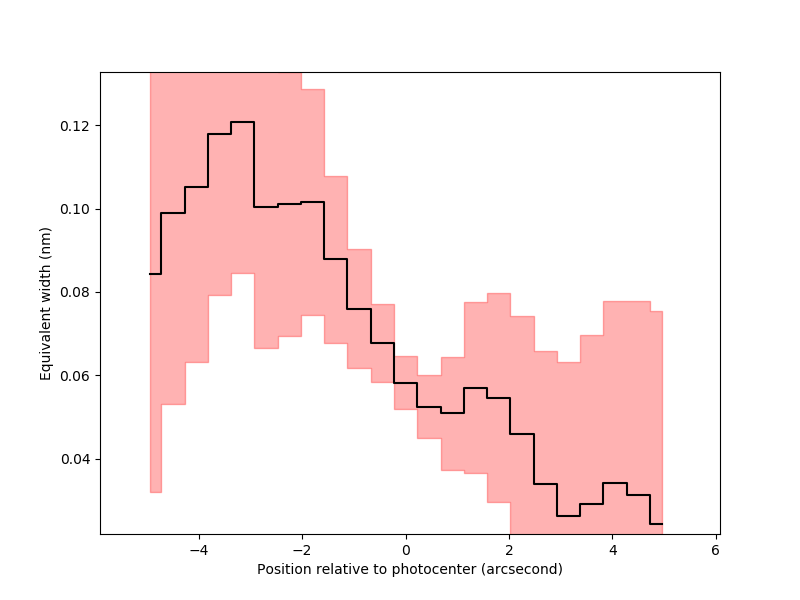}
   \caption{\label{1118_ew} Equivalent width of the 1118 nm absorption feature as a function of the position along the slit, measured with 0.45" bins. The red region represents 3-sigma uncertainties}
\end{figure}

\begin{figure}[ht]
   \includegraphics[width=0.44\textwidth]{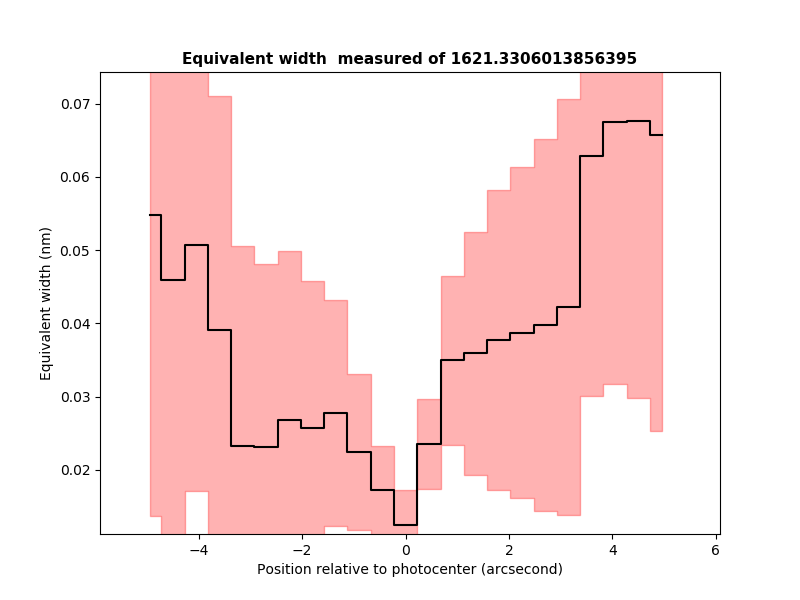}
   \caption{\label{1620_ew} Equivalent width of the 1620 nm absorption feature as a function of the position along the slit, measured with 0.45" bins. The red region represents 3-sigma uncertainties.}
\end{figure}

Two pairs of absorption features are weakly but clearly detected in our spectrum. The first contains lines at 1111~nm and 1118~nm, and the second lies at 1590~nm and 1620~nm. For each pair we present the spatial distribution of the equivalent width of the most prominent line (1118~nm and 1620~nm) in Figs.~\ref{1118_ew} and \ref{1620_ew}.

The equivalent width of the 1118 nm line exhibits a constant increase from north to south, while the equivalent width of the 1620 nm line is at its minimum in the central region and increases in both directions away from the the CE location. The signal-to-noise ratio of the 1590 nm and 1620 nm lines \textbf{is low} and the  1111 nm and 1118 nm overlap, therefore  no Doppler shift can be measured on those features.
\section{Discussion}
\label{discussion}
\subsection*{A young stellar cluster at the heart of NGC 1068?}

Our results confirm the significant contribution of stellar light at the centre of NGC 1068: except in the very central region where it accounts for $\sim$40$\%$ of the YJH flux, the stellar light largely dominates the spectrum in all the three bands of our observation (YJH). This large contribution is in good agreement with previous findings such as reported by \cite{Origlia1993}, who concluded from absorption features that stellar light accounts for $\sim$70$\%$ of the flux in the H band in the $4.4"\times4.4"$ central region, by \cite{Storchi-Bergmann2012}, who concluded from spectral synthesis in H and K bands that the stellar population dominates the spectrum in the inner 180 pc - 2", or most recently, by \cite{Rouan2019}, who revealed a young stellar cusp in the most central region of NGC 1068 based on high angular resolution imaging with SPHERE. Our analysis provided a well-defined spatial profile for the stellar component (Fig. \ref{plot_fluxs}), revealing a central cluster with a 2" radial extent (whose centre is co-located with the maximum of the 830~K emission) that is over-imposed on the very young stellar background. \\

We have also been able to compare the observed continuum of this cluster to synthetic spectra of an SSP at different ages and found that an SSP of 120 Myr provides the best fit, suggesting a recent starburst phase in this region.  \cite{Davies2007} also found indications of a starburst phase, but \textbf{longer ago} (250 Myr), which is in agreement with \cite{Martins2010a}, whose stellar population synthesis indicates an intermediate-age population together with some small contribution of young stars. Finally, \cite{Storchi-Bergmann2012} found two populations: one of 30 Myr, and the another of 300 Myr, similar to \cite{Davies2007}. Our age estimate for the central cluster then seems to be lower than those of the cited studies. However, considering that our method is mainly sensitive to the temperature of the stars, a slight excess of hot stars can easily be interpreted as a more recent star formation when the population with an SSP is modelled. We consider this the best justification for the observed discrepancy but do not conclude on the origin of these hot stars, which might  either be physically present in the central cluster or be a residual from the background subtraction.
\textbf{Despite} this difference in the age estimate, the picture we presented here of a young stellar cluster surrounding the CE that is over-imposed on a background composed of very hot stars is consistent with previous observations. \\
Additional indicators of stellar population available in our data are the absorption features at 1111, 1118, 1590, and 1620 nm. The 1590 nm feature is mostly caused by Si I, and the 1620 nm feature is a CO bandhead. Both trace cold stars (G, K, and early-M stars), as explained in \cite{Origlia1993}. We find no variation in the ratio of these two lines along the slit and conclude that they trace a population with a uniform composition. The equivalent width of these features (Fig. \ref{1620_ew}) is consistent with the scheme we also deduced from our continuum analysis: an extended stellar population diluted with other components in the central region.
On the other hand, as suggested by \cite{Riffel2006} and \cite{Maraston2005}, the 1111 and 1118 nm features can be associated with CN absorption from $\sim$1 Gyr thermal-pulsing AGB stars. However, the equivalent width along the slit of these features (Fig. \ref{1118_ew}) is quite different from the other stellar lines. It exhibits an almost constant increase from north to south without the expected dilution by the nucleus in the central region. This clearly indicates that these lines originate from a structure that is different from the structure that is probed with the Si I and CO absorption features. They might trace an older stellar population, possibly the population that was detected by \cite{Davies2007}. The lack of dilution in the central region might also suggest that these lines arise from foreground interstellar material. However, we found no trace of interstellar absorption lines at these wavelengths in dedicated catalogues, which supports the hypothesis of a stellar origin even if the spatial variation of the equivalent width remains unexplained.\\

\subsection*{Dust}
The observed $830$ K blackbody component very likely traces the obscuring dusty structure that is predicted by the unified model. The temperature is consistent with previous observations at longer wavelengths \citep{Gratadour2003,Exposito2011,Storchi-Bergmann2012}. The emission of this component comes from an unresolved region (< 0.4") at the location of the photocenter. The unresolved region is responsible for more than $50\ \%$ of the flux. No trace from the nodules observed at longer wavelength \citep{Gratadour2005,Exposito2011} were detected in our observation, probably because our spatial resolution is lower. \\

Spectroscopic observations of Seyfert 2 galaxies often exhibit a minimum in the NIR continuum emission, whose location varies from 1.1 to 1.4 $\mu m$ \citep{Riffel2006}. In our analysis, we conclude that the position of this minimum is determined by the relative significance of stellar and dust components. \\
\subsection*{Contribution of scattered light to the continuum emission}
Several studies pointed out that scattered light from the CE may significantly contribute to the NIR continuum: \cite{Crenshaw2000} concluded that more than $69\%$ of the UV-to-NIR flux from the central region comes from the hidden nucleus, while \cite{Martins2010a} found a $25\%$ contribution to the $0.8-2.4\ \mu m$ continuum in the centre with peaks at 100-150 pc in the north and south directions. The latter observations are consistent with scattered light detected with polarimetric imaging in the nuclear and circumnuclear regions by \cite{Gratadour2015}. To measure the possible contribution of a Seyfert 1 nucleus to our spectrum, we introduced a component with a flat spectrum in our model. Such a flat spectrum was observed in a Seyfert 1 galaxy by \cite{Riffel2006}. A significant contribution was clearly detected in the unresolved central region where hot dust prevails as well as a significant contribution from 2" to 4" south of the CE. This is very well explained in polarised intensity maps from \cite{Gratadour2015}: A slit centred on the CE and orientated with $PA = 12$\degr \ intercepts two polarised regions: the first in the very centre, and the second 3" south. Moreover, our flux estimates are compatible with these maps. We find a $\sim 5 \% $ contribution of scattered light in the H band in the central region and a $\sim 15 \% $ in the southern region \citep[compared to $\sim 6 \% $ in the central region and $\sim 10 \% $ with peaks at $13 \%$ in the southern region for][]{Gratadour2015}. The consistency between these results confirms the interpretation of polarised light as scattered light from the CE and our ability to detect it with $\lambda^0$ power law. 
The slope of the input scattered light \citep[from $\lambda^{-1.5}$ , as suggested by ][to $\lambda^0$ as presented in this paper]{Crenshaw2000} has very little effect on our results because it is very different from the stellar and hot-dust component slopes ($\sim \lambda^{-3}$ and $\sim \lambda^{5}$, respectively) in any case. It is then always detected at the same location with a similar contribution to the flux. \\

\subsection*{Ionised gas}
\begin{figure}[ht]
   \includegraphics[width=0.44\textwidth]{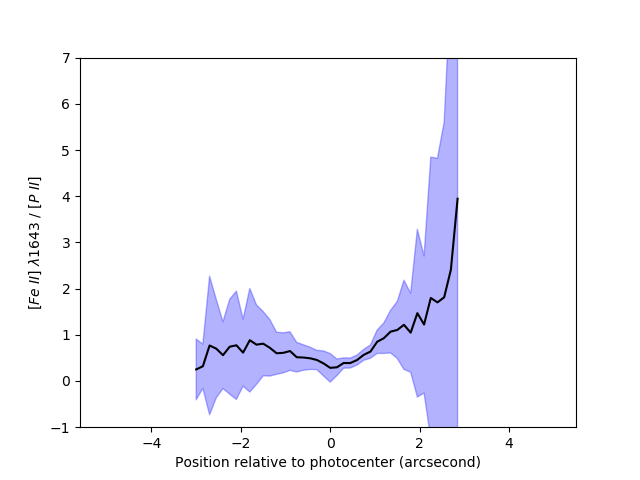}
   \caption{\label{ratio} Spatial distribution of the $\left[\mathrm{Fe~II}\right]\mathrm{~1643}\ /\ \left[\mathrm{P~II}\right] $ ratio.}
\end{figure}

Except for the [Fe~II] lines, all the emission features we detected in our observation follow a similar behaviour in their spatial distribution and their Doppler shift. We found that their position of maximum emission is located very near the photocenter, from the exact same position (He~I, $\left[\mathrm{P~II}\right] $, $\left[\mathrm{S~IX}\right] $, Pa~$\beta$) to 0.3" north ($\left[\mathrm{S~II}\right] $). The spatial extent of these emission lines extends to 4" (detection limit), with FWHM $\sim$1", and presents an asymmetry with a significant emission excess at the northern side of the slit. This asymmetric behaviour can first be explained by orientation effects. \cite{Das2006} have concluded that the NLR is inclined toward us ($i=5^o$), and its northern part lies slightly closer to us than the southern part. This orientation may favour the detection in the north of some regions whose southern counterparts are hidden from us. Secondly, this emission line excess might alternatively arise from the nodules that have been detected in \cite{Gratadour2006} and have previously been associated with emission lines in the K band in \cite{Exposito2011}. Observations with higher angular resolution are required to distinguish the role of these two possible causes in the asymmetric emission line distribution.     \\
Supporting the first scenario, a significant Doppler shift is detected for these lines that depends on the position on the slit. It most probably traces a radial outflow, as observed in previous studies \citep{Cecil2002,Das2006}. We can observe in He I, the line with the clearest signal, \textbf{a speed that constantly} increases up to 1.5" (120 pc), followed by a quick drop back to rest speed at 2" (160 pc).  The drop in speed seems correlated with a structure detected in NIR polarimetric images \citep{Gratadour2015}. The spatial coincidence between the deceleration of the ionised gas and the scattering material indicates that the deceleration is probably caused by the interaction of the NLR with this structure. \\
Among the emission lines, the coronal lines (lines with ionisation energy > 100 eV, $\left[\mathrm{S~IX}\right]$ and  $\left[\mathrm{Si~X}\right]$ in our sample) seem to exhibit a narrower spatial profile. This observation is consistent with the scheme of photoionisation from the strong nuclear UV-X continuum, whose intensity quickly decreases with distance to the nucleus.\\
In comparison, the $\left[\mathrm{Fe~II}\right]$ 1643 emission line reveals unusual features. First, it is detected in a much more extended region (i.e. twice the size of the emitting region of other lines). In the scenario involving photoionisation by the CE, this can be explained by its low ionisation potential that allows ionisation at greater distances from the source. However, it also appears that the peak emission of this emission line has a peak emission that does not coincide with the photocenter as other lines do. This probably indicates a different structure, and possibly a different ionisation mechanism. In order to test the possibility of shock-induced ionisation, we studied the $\left[\mathrm{Fe~II}\right]$ 1643$\ /\ \left[\mathrm{P~II}\right] $ ratio, which is assumed to be sensitive to the ionisation mechanism: values close to 1 indicate photoionisation, while higher values ($\sim 10$) indicate shocks. This diagnostic tool comes from the fact that iron is assumed to be locked in dust grains and that a strong $\left[\mathrm{Fe~II}\right]$ emission indicates that dust grains must have been destroyed, shocks being the most plausible explanation. A similar diagnosis was performed by \cite{Oliva2001} and  \cite{Hashimoto2011} with the $\left[\mathrm{Fe~II}\right]$~1257 line instead. The latter has the advantage of being close to $\left[\mathrm{P~II}\right] $ and the diagnosis is then less sensitive to extinction. However, in our case, the measure of $\left[\mathrm{Fe~II}\right]$ 1257 is too inaccurate because of its proximity with $\left[\mathrm{S~IX}\right], $ and we preferred to use $\left[\mathrm{Fe~II}\right]$~1643, whose ratio with $\left[\mathrm{Fe~II}\right]$~1257  is fixed at 1.35 in the classical case B scenario from \cite{Osterbrock1989}. As shown in Fig. \ref{ratio}, this ratio remains well below 10 everywhere it can be measured, indicating that shocks do not play a significant role in the ionisation of $\left[\mathrm{Fe~II}\right]$~1643.\\

In addition to its spatial distribution, the $\left[\mathrm{Fe~II}\right]$ 1643 emission line also presents a very different velocity profile. A much stronger Doppler redshift than the one from the outflow ($v\sim600\ km.s^{-1}$ versus $v\sim200\ km.s^{-1}$, see Fig. \ref{fe_ii_speed} and Fig. \ref{he_i_speed}, respectively) affects the line in the northern region. This difference confirms the previous suggestion that this emission arises from a different structure than the other emission lines, which possibly traces a foreground cloud of gas that falls onto the CE.

\section{Conclusion}

In order to understand the physical processes that occur in the central region of NGC 1068, we performed a long-slit spectroscopy analysis of the nucleus in YJH bands (0.95 - 1.65 $\mu m$) at sub-arcsecond angular resolution (0.35"). We presented a method to analyse the continuum emission and showed that the emission can be decomposed into four components: a young stellar population (120 Myr), hot dust (830~K), scattered light from the hidden Seyfert 1 nucleus, and a very hot stellar background. In the very central region hot dust is the main contributor to the flux (more than 50\%) but scattered light is also significantly detected ($\sim$ 10$\%$). This level of scattered light is consistent with polarimetric measurements in the H band. Elsewhere, the stellar content is the main source of continuum emission, with an extended very hot stellar population that covers the entire slit and a cuspy 120 Myr stellar population around the photocenter. The good agreement between our results and previous studies on the same object confirms the ability of our method to determine the spatial distribution of these components. It has the advantage to be suitable for medium-resolution spectra (R=350) because it does not use absorption lines to fit the stellar population. For the same reason, it is less sensitive than other methods to the signal-to-noise ratio of the spectrum. Finally, it allows basic detection of the scattered light without the use of polarimetric measurements. \\
Absorption features at 1.59 and 1.62 $\mu m$ that tarce cold stars are detected in an extended region, supporting the previous conclusion that stellar content is the main contributor to the observed flux.\\
Many emission lines are detected that trace the NLR of NGC 1068. Almost all these emission lines emit most strongly in the central 0.5", which supports the hypothesis of photoionisation from the CE. Moreover, our analysis of the $\left[\mathrm{Fe~II}\right]\mathrm{~1643}\ /\ \left[\mathrm{P~II}\right] $ does not support a shock-induced ionisation scenario. A Doppler shift was measured in several of these lines ($\left[\mathrm{S~II}\right]$, He~I, $\left[\mathrm{P~II}\right] $, and Pa~$\beta$ , leading to the conclusion that they trace the outflow that originates from around the nucleus, whose northern part moves towards the observer, while southern part moves away.\\
The $\left[\mathrm{Fe~II}\right]$ emission line features are quite different.  Their spatial distribution is more extended, their position of maximum emission is offset north of the nucleus, and they present a very different Doppler shift. We conclude that they arise from a distinct structure that possibly traces an inflow to the CE.  \\  
We showed the potential of the YJH bands for the study of Seyfert 2 nuclei. A full AO correction could add valuable information, in particular, for the study of the very central region. Moreover, other slit orientations could help to investigate locations of interest, in particular, the small nodules that surround the nucleus and the polarised spots that we detected in the circumnuclear region.    

\bibliographystyle{aa}
\bibliography{biblio}

\begin{thebibliography}{44}
\expandafter\ifx\csname natexlab\endcsname\relax\def\natexlab#1{#1}\fi

\bibitem[{{Antonucci}(1993)}]{Antonucci1993}
{Antonucci}, R. 1993, \araa, 31, 473

\bibitem[{{Antonucci} \& {Miller}(1985)}]{Antonucci1985}
{Antonucci}, R.~R.~J. \& {Miller}, J.~S. 1985, \apj, 297, 621

\bibitem[{{Beuzit} {et~al.}(2008){Beuzit}, {Feldt}, {Dohlen}, {Mouillet},
  {Puget}, {Wildi}, {Abe}, {Antichi}, {Baruffolo}, {Baudoz}, {Boccaletti},
  {Carbillet}, {Charton}, {Claudi}, {Downing}, {Fabron}, {Feautrier},
  {Fedrigo}, {Fusco}, {Gach}, {Gratton}, {Henning}, {Hubin}, {Joos}, {Kasper},
  {Langlois}, {Lenzen}, {Moutou}, {Pavlov}, {Petit}, {Pragt}, {Rabou}, {Rigal},
  {Roelfsema}, {Rousset}, {Saisse}, {Schmid}, {Stadler}, {Thalmann}, {Turatto},
  {Udry}, {Vakili}, \& {Waters}}]{Beuzit2008}
{Beuzit}, J.-L., {Feldt}, M., {Dohlen}, K., {et~al.} 2008, in \procspie, Vol.
  7014, Ground-based and Airborne Instrumentation for Astronomy II, 701418

\bibitem[{{Bock} {et~al.}(2000){Bock}, {Neugebauer}, {Matthews}, {Soifer},
  {Becklin}, {Ressler}, {Marsh}, {Werner}, {Egami}, \& {Blandford}}]{Bock2000}
{Bock}, J.~J., {Neugebauer}, G., {Matthews}, K., {et~al.} 2000, \aj, 120, 2904

\bibitem[{{Cecil} {et~al.}(2002){Cecil}, {Dopita}, {Groves}, {Wilson},
  {Ferruit}, {P{\'e}contal}, \& {Binette}}]{Cecil2002}
{Cecil}, G., {Dopita}, M.~A., {Groves}, B., {et~al.} 2002, \apj, 568, 627

\bibitem[{{Crenshaw} \& {Kraemer}(2000)}]{Crenshaw2000}
{Crenshaw}, D.~M. \& {Kraemer}, S.~B. 2000, \apj, 532, 247

\bibitem[{{Cutri} {et~al.}(2003){Cutri}, {Skrutskie}, {van Dyk}, {Beichman},
  {Carpenter}, {Chester}, {Cambresy}, {Evans}, {Fowler}, {Gizis}, {Howard},
  {Huchra}, {Jarrett}, {Kopan}, {Kirkpatrick}, {Light}, {Marsh}, {McCallon},
  {Schneider}, {Stiening}, {Sykes}, {Weinberg}, {Wheaton}, {Wheelock}, \&
  {Zacarias}}]{Cutri2003}
{Cutri}, R.~M., {Skrutskie}, M.~F., {van Dyk}, S., {et~al.} 2003, VizieR Online
  Data Catalog, II/246

\bibitem[{{Das} {et~al.}(2006){Das}, {Crenshaw}, {Kraemer}, \& {Deo}}]{Das2006}
{Das}, V., {Crenshaw}, D.~M., {Kraemer}, S.~B., \& {Deo}, R.~P. 2006, \aj, 132,
  620

\bibitem[{{Davies} {et~al.}(2007{\natexlab{a}}){Davies}, {Genzel}, {Tacconi},
  {M{\"u}ller S{\'a}nchez}, \& {Sternberg}}]{Davies2007}
{Davies}, R., {Genzel}, R., {Tacconi}, L., {M{\"u}ller S{\'a}nchez}, F., \&
  {Sternberg}, A. 2007{\natexlab{a}}, in The Central Engine of Active Galactic
  Nuclei, Vol. 373, 639

\bibitem[{{Davies} {et~al.}(2007{\natexlab{b}}){Davies}, {M{\"u}ller
  S{\'a}nchez}, {Genzel}, {Tacconi}, {Hicks}, {Friedrich}, \&
  {Sternberg}}]{Davies2007a}
{Davies}, R.~I., {M{\"u}ller S{\'a}nchez}, F., {Genzel}, R., {et~al.}
  2007{\natexlab{b}}, \apj, 671, 1388

\bibitem[{{Dopita} \& {Sutherland}(1996)}]{Dopita1996}
{Dopita}, M.~A. \& {Sutherland}, R.~S. 1996, \apjs, 102, 161

\bibitem[{{Exposito} {et~al.}(2011){Exposito}, {Gratadour}, {Cl{\'e}net}, \&
  {Rouan}}]{Exposito2011}
{Exposito}, J., {Gratadour}, D., {Cl{\'e}net}, Y., \& {Rouan}, D. 2011, \aap,
  533, A63

\bibitem[{{Gallimore} {et~al.}(1996){Gallimore}, {Baum}, {O'Dea}, \&
  {Pedlar}}]{Gallimore1996}
{Gallimore}, J.~F., {Baum}, S.~A., {O'Dea}, C.~P., \& {Pedlar}, A. 1996, \apj,
  458, 136

\bibitem[{{Gratadour} {et~al.}(2003){Gratadour}, {Cl{\'e}net}, {Rouan}, {Lai},
  \& {Forveille}}]{Gratadour2003}
{Gratadour}, D., {Cl{\'e}net}, Y., {Rouan}, D., {Lai}, O., \& {Forveille}, T.
  2003, \aap, 411, 335

\bibitem[{{Gratadour} {et~al.}(2015){Gratadour}, {Rouan}, {Grosset},
  {Boccaletti}, \& {Cl{\'e}net}}]{Gratadour2015}
{Gratadour}, D., {Rouan}, D., {Grosset}, L., {Boccaletti}, A., \& {Cl{\'e}net},
  Y. 2015, \aap, 581, L8

\bibitem[{{Gratadour} {et~al.}(2006){Gratadour}, {Rouan}, {Mugnier}, {Fusco},
  {Cl{\'e}net}, {Gendron}, \& {Lacombe}}]{Gratadour2006}
{Gratadour}, D., {Rouan}, D., {Mugnier}, L.~M., {et~al.} 2006, \aap, 446, 813

\bibitem[{{Gratadour} {et~al.}(2005){Gratadour}, {Rouan, D.}, {Boccaletti, A.},
  {Riaud, P.}, \& {Cl\'enet, Y.}}]{Gratadour2005}
{Gratadour}, D., {Rouan, D.}, {Boccaletti, A.}, {Riaud, P.}, \& {Cl\'enet, Y.}
  2005, A\&A, 429, 433

\bibitem[{{Hashimoto} {et~al.}(2011){Hashimoto}, {Nagao}, {Yanagisawa},
  {Matsuoka}, \& {Araki}}]{Hashimoto2011}
{Hashimoto}, T., {Nagao}, T., {Yanagisawa}, K., {Matsuoka}, K., \& {Araki}, N.
  2011, Publications of the Astronomical Society of Japan, 63, L7

\bibitem[{{Hickox} \& {Alexander}(2018)}]{Hickox2018}
{Hickox}, R.~C. \& {Alexander}, D.~M. 2018, \araa, 56, 625

\bibitem[{{Jaffe} {et~al.}(2004){Jaffe}, {Meisenheimer}, {R{\"o}ttgering},
  {Leinert}, {Richichi}, {Chesneau}, {Fraix-Burnet}, {Glazenborg-Kluttig},
  {Granato}, {Graser}, {Heijligers}, {K{\"o}hler}, {Malbet}, {Miley},
  {Paresce}, {Pel}, {Perrin}, {Przygodda}, {Schoeller}, {Sol}, {Waters},
  {Weigelt}, {Woillez}, \& {de Zeeuw}}]{Jaffe2004}
{Jaffe}, W., {Meisenheimer}, K., {R{\"o}ttgering}, H.~J.~A., {et~al.} 2004,
  \nat, 429, 47

\bibitem[{{Kotulla} {et~al.}(2009){Kotulla}, {Fritze}, {Weilbacher}, \&
  {Anders}}]{Kotulla2009}
{Kotulla}, R., {Fritze}, U., {Weilbacher}, P., \& {Anders}, P. 2009, \mnras,
  396, 462

\bibitem[{{Kraemer} \& {Crenshaw}(2000)}]{Kraemer2000}
{Kraemer}, S.~B. \& {Crenshaw}, D.~M. 2000, \apj, 544, 763

\bibitem[{{Lodato} \& {Bertin}(2003)}]{Lodato2003}
{Lodato}, G. \& {Bertin}, G. 2003, \aap, 398, 517

\bibitem[{{L{\'o}pez-Gonzaga} {et~al.}(2014){L{\'o}pez-Gonzaga}, {Jaffe},
  {Burtscher}, {Tristram}, \& {Meisenheimer}}]{Lopez-Gonzaga2014}
{L{\'o}pez-Gonzaga}, N., {Jaffe}, W., {Burtscher}, L., {Tristram}, K.~R.~W., \&
  {Meisenheimer}, K. 2014, \aap, 565, A71

\bibitem[{{Macchetto} {et~al.}(1994){Macchetto}, {Capetti}, {Sparks}, {Axon},
  \& {Boksenberg}}]{Macchetto1994}
{Macchetto}, F., {Capetti}, A., {Sparks}, W.~B., {Axon}, D.~J., \&
  {Boksenberg}, A. 1994, \apjl, 435, L15

\bibitem[{{Maraston}(2005)}]{Maraston2005}
{Maraston}, C. 2005, \mnras, 362, 799

\bibitem[{{Martins} {et~al.}(2010{\natexlab{a}}){Martins}, {Riffel},
  {Rodr{\'\i}guez- Ardila}, {Gruenwald}, \& {de Souza}}]{Martins2010}
{Martins}, L.~P., {Riffel}, R., {Rodr{\'\i}guez- Ardila}, A., {Gruenwald}, R.,
  \& {de Souza}, R. 2010{\natexlab{a}}, \mnras, 406, 2185

\bibitem[{{Martins} {et~al.}(2010{\natexlab{b}}){Martins},
  {Rodr{\'\i}guez-Ardila}, {de Souza}, \& {Gruenwald}}]{Martins2010a}
{Martins}, L.~P., {Rodr{\'\i}guez-Ardila}, A., {de Souza}, R., \& {Gruenwald},
  R. 2010{\natexlab{b}}, \mnras, 406, 2168

\bibitem[{Mor \& Netzer(2012)}]{Mor2012}
Mor, R. \& Netzer, H. 2012, Monthly Notices of the Royal Astronomical Society,
  420, 526

\bibitem[{Nazarova(1996)}]{Nazarov1996}
Nazarova, L. 1996, Vistas in Astronomy, 40, 35 , proceedings of the Oxford
  Torus Workshop

\bibitem[{{Oliva} {et~al.}(2001){Oliva}, {Marconi}, {Maiolino}, {Testi},
  {Mannucci}, {Ghinassi}, {Licandro}, {Origlia}, {Baffa}, {Checcucci},
  {Comoretto}, {Gavryussev}, {Gennari}, {Giani}, {Hunt}, {Lisi}, {Lorenzetti},
  {Marcucci}, {Miglietta}, {Sozzi}, {Stefanini}, \& {Vitali}}]{Oliva2001}
{Oliva}, E., {Marconi}, A., {Maiolino}, R., {et~al.} 2001, \aap, 369, L5

\bibitem[{{Origlia} {et~al.}(1993){Origlia}, {Moorwood}, \&
  {Oliva}}]{Origlia1993}
{Origlia}, L., {Moorwood}, A.~F.~M., \& {Oliva}, E. 1993, \aap, 280, 536

\bibitem[{{Osterbrock}(1989)}]{Osterbrock1989}
{Osterbrock}, D.~E. 1989, {Astrophysics of gaseous nebulae and active galactic
  nuclei}

\bibitem[{{Pickles}(1998)}]{Pickles1998}
{Pickles}, A.~J. 1998, \pasp, 110, 863

\bibitem[{{Pier} {et~al.}(1994){Pier}, {Antonucci}, {Hurt}, {Kriss}, \&
  {Krolik}}]{Pier1994}
{Pier}, E.~A., {Antonucci}, R., {Hurt}, T., {Kriss}, G., \& {Krolik}, J. 1994,
  \apj, 428, 124

\bibitem[{{Poncelet} {et~al.}(2006){Poncelet}, {Perrin}, \&
  {Sol}}]{Poncelet2006}
{Poncelet}, A., {Perrin}, G., \& {Sol}, H. 2006, \aap, 450, 483

\bibitem[{{Rees}(1984)}]{Rees1984}
{Rees}, M.~J. 1984, \araa, 22, 471

\bibitem[{{Riffel} {et~al.}(2006){Riffel}, {Rodr{\'{\i}}guez-Ardila}, \&
  {Pastoriza}}]{Riffel2006}
{Riffel}, R., {Rodr{\'{\i}}guez-Ardila}, A., \& {Pastoriza}, M.~G. 2006, \aap,
  457, 61

\bibitem[{{Riffel} {et~al.}(2014){Riffel}, {Vale}, {Storchi-Bergmann}, \&
  {McGregor}}]{Riffel2014}
{Riffel}, R.~A., {Vale}, T.~B., {Storchi-Bergmann}, T., \& {McGregor}, P.~J.
  2014, \mnras, 442, 656

\bibitem[{{Romeo} \& {Fathi}(2016)}]{Romeo2016}
{Romeo}, A.~B. \& {Fathi}, K. 2016, \mnras, 460, 2360

\bibitem[{{Rouan} {et~al.}(2019){Rouan}, {Grosset}, \& {Gratadour}}]{Rouan2019}
{Rouan}, D., {Grosset}, L., \& {Gratadour}, D. 2019, \aap, 625, A100

\bibitem[{{Storchi-Bergmann} {et~al.}(2012){Storchi-Bergmann}, {Riffel},
  {Riffel}, {Diniz}, {Borges Vale}, \& {McGregor}}]{Storchi-Bergmann2012}
{Storchi-Bergmann}, T., {Riffel}, R.~A., {Riffel}, R., {et~al.} 2012, \apj,
  755, 87

\bibitem[{{Vigan} {et~al.}(2008){Vigan}, {Langlois}, {Moutou}, \&
  {Dohlen}}]{Vigan2008}
{Vigan}, A., {Langlois}, M., {Moutou}, C., \& {Dohlen}, K. 2008, \aap, 489,
  1345

\bibitem[{{Wilson} \& {Ulvestad}(1983)}]{Wilson1983}
{Wilson}, A.~S. \& {Ulvestad}, J.~S. 1983, \apj, 275, 8

\end{thebibliography}

\begin{appendix}
\section{Line profiles}

\begingroup
\centering
\begin{figure*}
\centering
\caption{Spatial profile of absorption lines}
\subfigure[1111 nm]{\includegraphics[width=0.3\textwidth]{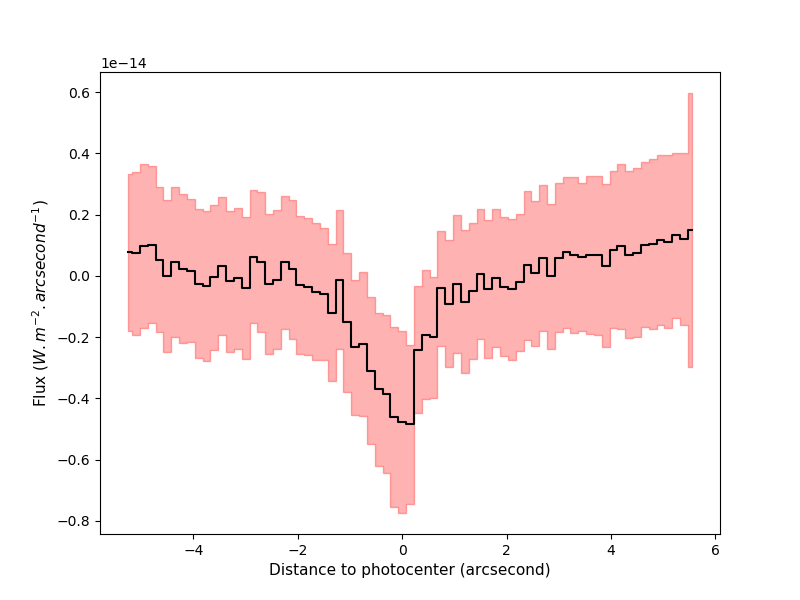}}\label{1111_prof_annexe}
\subfigure[1590 nm]{\includegraphics[width=0.3\textwidth]{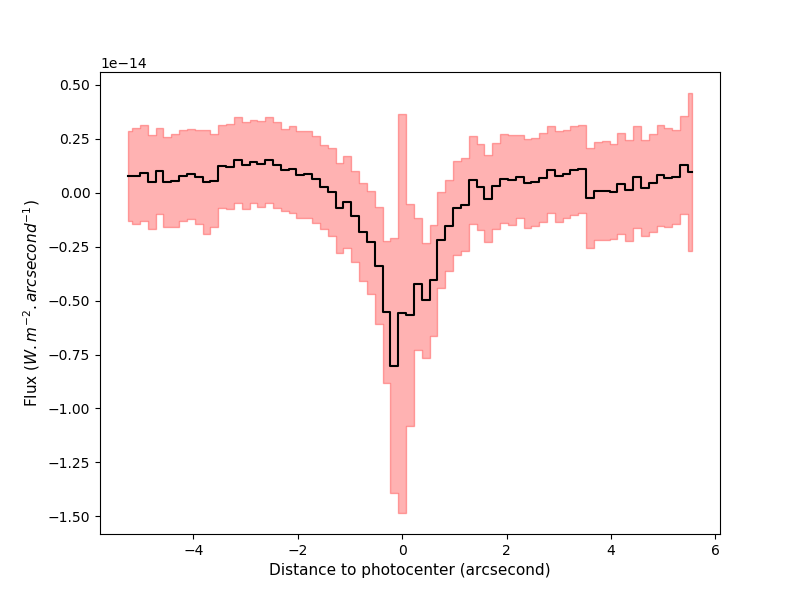}}\label{1590_prof_annexe}
\label{prof_abs_annexe}
\end{figure*}
\endgroup

\begingroup
\centering
\begin{figure*}
\centering
\caption{Spatial profile of emission lines}
\subfigure[He~II]{\includegraphics[width=0.3\textwidth]{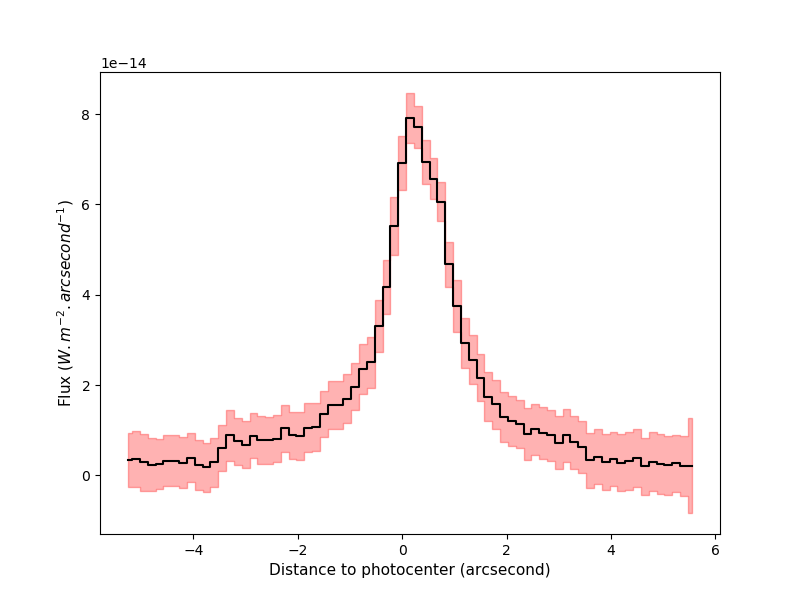}}\label{he_ii_prof_annexe}
\subfigure[\sii]{\includegraphics[width=0.3\textwidth]{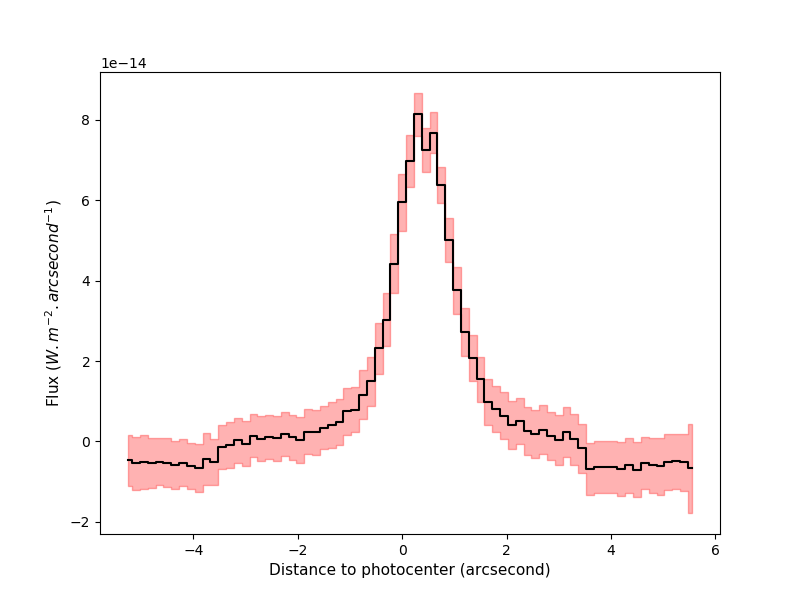}}\label{s_ii_prof_annexe}
\subfigure[\pii]{\includegraphics[width=0.3\textwidth]{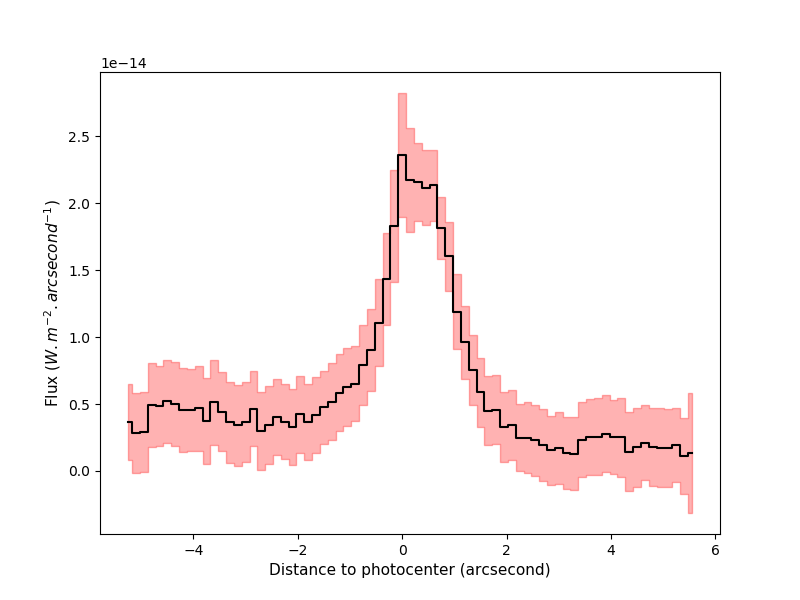}}\label{p_ii_prof_annexe}
\subfigure[\pagam]{\includegraphics[width=0.3\textwidth]{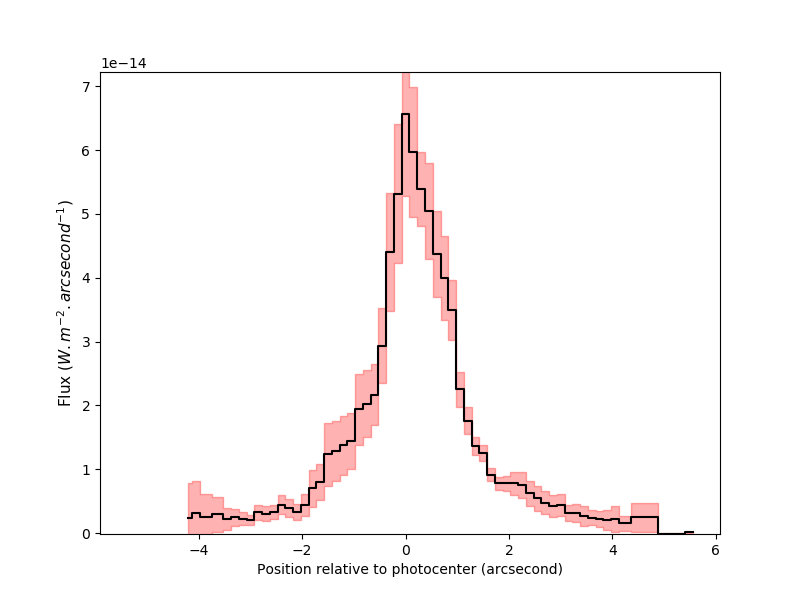}}\label{pagam_prof_annexe}
\subfigure[\sx]{\includegraphics[width=0.3\textwidth]{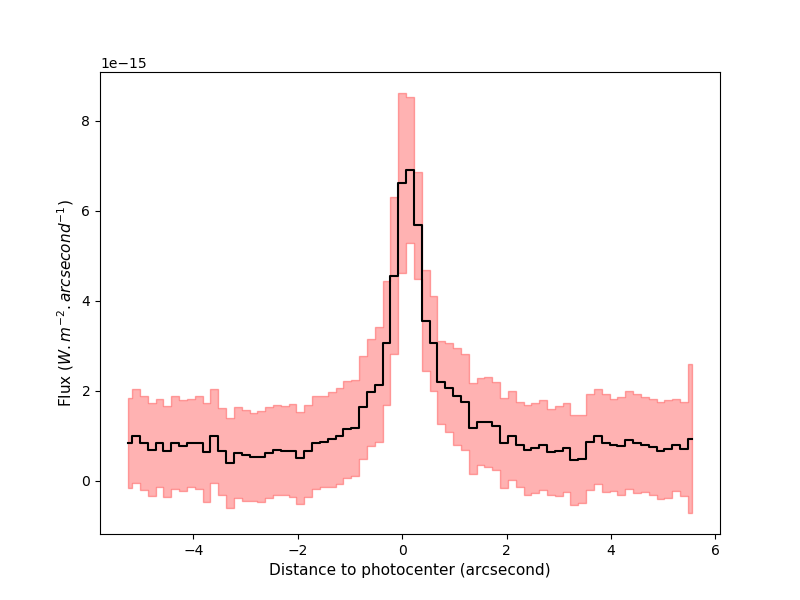}}\label{s_ix_prof_annexe}
\subfigure[\feiimilledeux]{\includegraphics[width=0.3\textwidth]{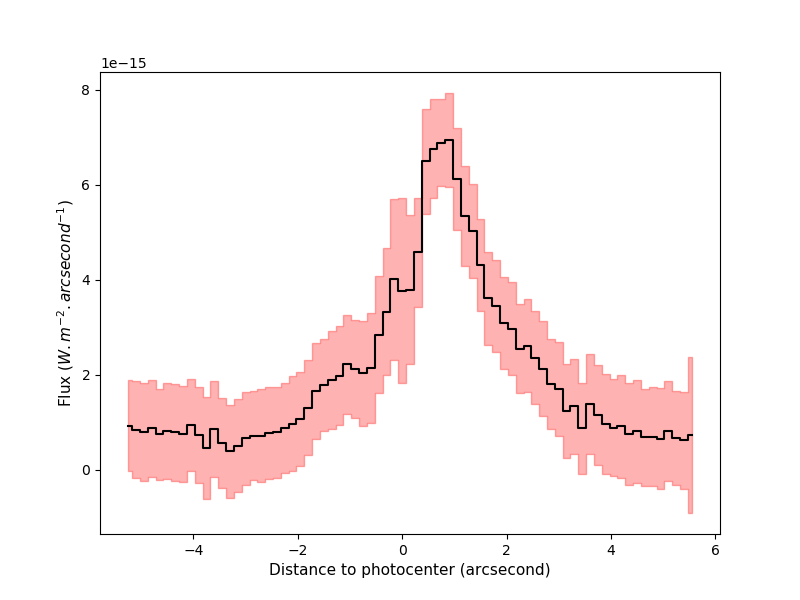}}\label{fe_ii_1257_prof_annexe}
\subfigure[\pabet]{\includegraphics[width=0.3\textwidth]{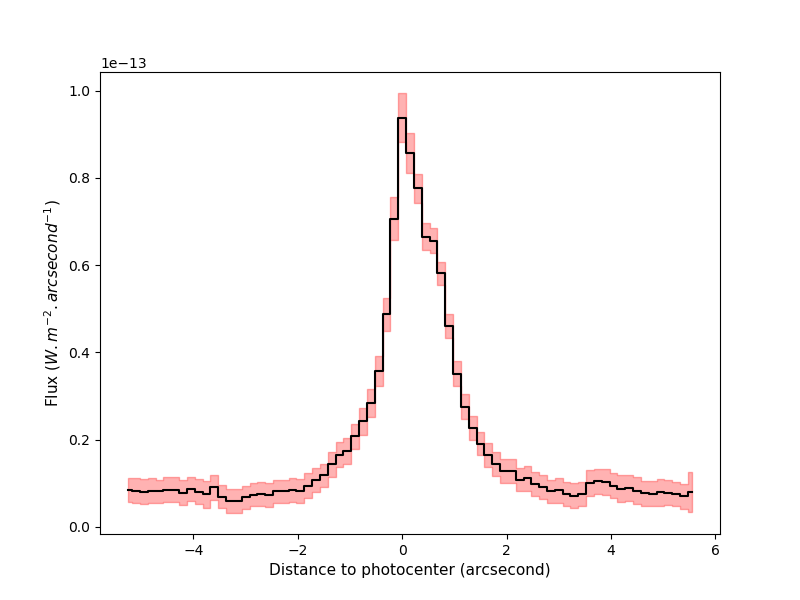}}\label{pabet_prof_annexe}
\subfigure[\six]{\includegraphics[width=0.3\textwidth]{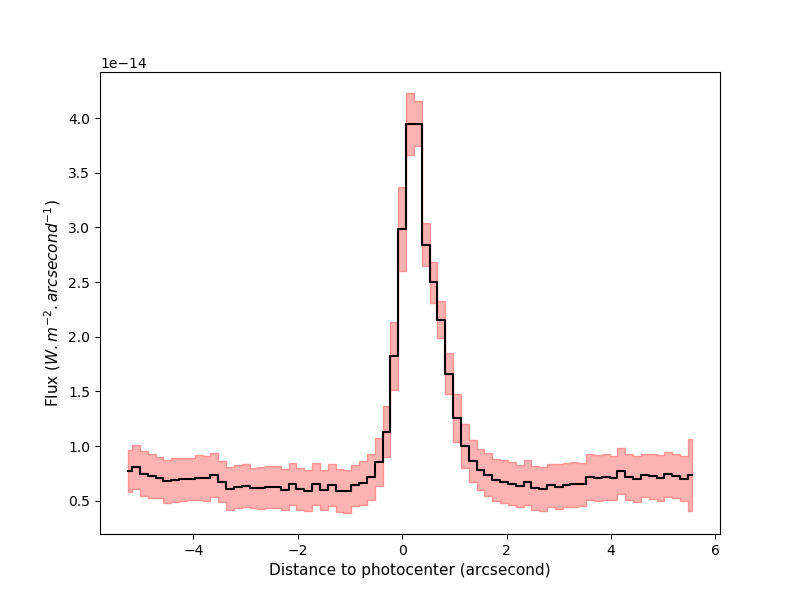}}\label{si_x_prof_annexe} 
\label{prof_em_annexe}
\end{figure*}
\endgroup

\section{Emission line Doppler shift}

\begingroup
\centering
\begin{figure*}
\centering
\caption{Doppler shift of emission lines}
\subfigure[\sii]{\includegraphics[width=0.3\textwidth]{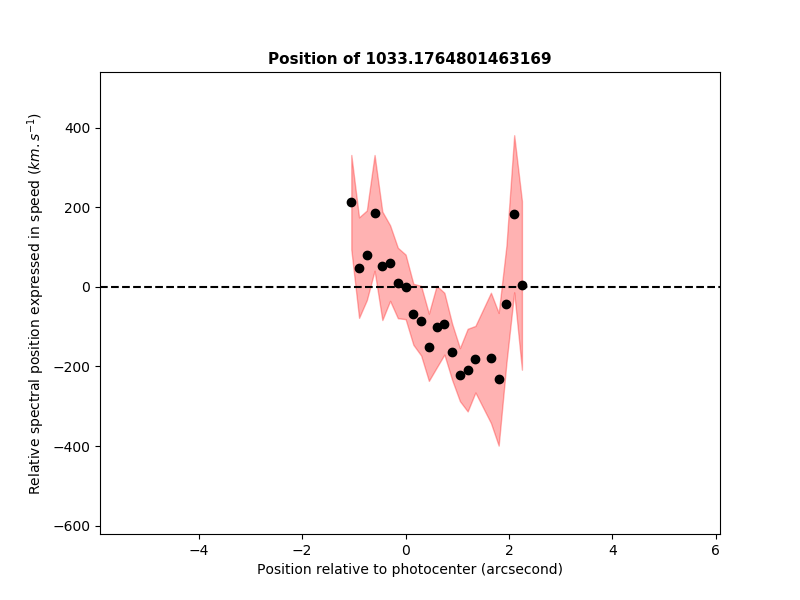}}\label{s_ii_speed_annexe}
\subfigure[\pii] {\includegraphics[width=0.3\textwidth]{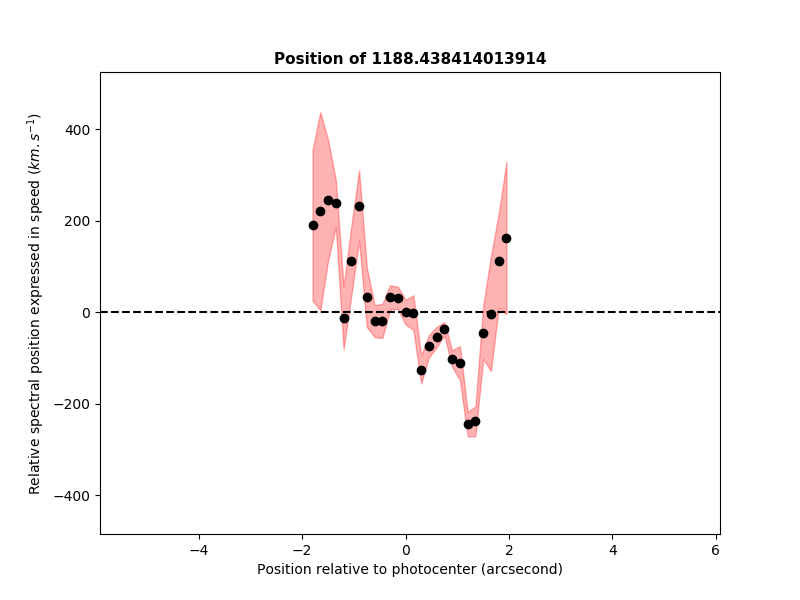}}\label{p_ii_speed_annexe}
\subfigure[\pagam]{\includegraphics[width=0.3\textwidth]{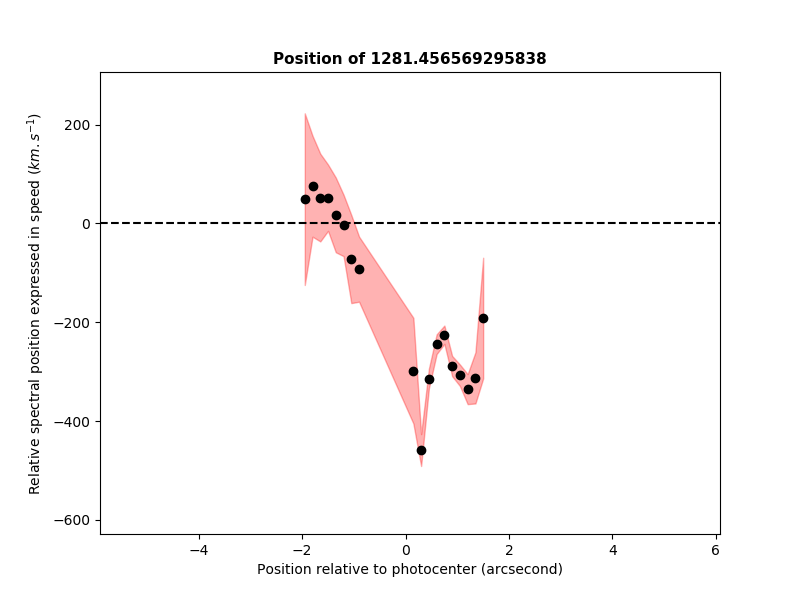}}\label{pagam_speed_annexe} \label{speed_annexe}
\end{figure*}

\endgroup

\end{appendix}
\end{document}